\documentclass[12pt]{iopart}
\usepackage{amssymb}
\usepackage{graphicx}

\def \beq {\begin{equation}}
\def \eeq {\end{equation}}
\def \ba {\begin{eqnarray}}
\def \ea {\end{eqnarray}}
\def \ban {\begin{eqnarray*}}
\def \ean {\end{eqnarray*}}
\def \r {\right}
\def \l {\left}

\newcommand{\ketbrad}[1]{|#1\rangle\!\langle #1|}
\newcommand{\ketbradt}[2]{|#1\rangle\!\langle #2|}

\newcommand{\mc}[1]{{\mathcal #1}}
\newcommand{\mean}[1]{\langle#1\rangle}

\renewcommand{\tr}[1]{\textrm{Tr}\left[#1\right]}
\def\ket#1{\left| #1\right>}
\def\bra#1{\left< #1\right|}

\begin{document}

\title{Bounds on quantum communication via Newtonian gravity}
\author{D. Kafri$^1$, G. Milburn$^{2,3}$, J. M. Taylor$^{1,4}$}
\address{$^1$ Joint Quantum Institute,
University of Maryland, College Park,  MD, USA}
\address{$^2$ Centre for Engineered Quantum Systems, School of Mathematics and Physics, The University of Queensland,  Australia.}
\address{$^3$ Kavli Institute for Theoretical Physics, University of California, Santa Barbara}
\address{$^4$ National Institute of Standards and Technology, Gaithersburg, MD, USA}
\date{\today}
\begin{abstract}
Newtonian gravity yields specific observable consequences, the most striking of which is the emergence of a $1/r^2$ force. In so far as communication can arise via such interactions between distant particles, we can ask what would be expected for a theory of gravity that only allows classical communication. Many heuristic suggestions for gravity-induced decoherence have this restriction implicitly or explicitly in their construction. Here we show that communication via a $1/r^2$ force  has a minimum noise induced in the system when the communication cannot convey quantum information, in a continuous time analogue to Bell's inequalities. Our derived noise bounds provide tight constraints from current experimental results on any theory of gravity that does not allow quantum communication.
\end{abstract}
\maketitle

Efforts to confirm the nature of gravity at a quantum mechanical level remain so far in the realm of theoretical exercise, with observational consequences of Planck-scale physics largely out of the reach of current and projected experiments.  However, recent progress on the theoretical limits of quantum theories have suggested~\cite{Karolyhazy,Penrose,Diosi,diosi2007,Pullin,Romero-Isart,Yang} that there may be effects due to gravity that occur in quantum systems that preclude a variety of potential approaches for reconciling quantum field theory with general relativity.  This has led to theories that question the need for a quantum theory of gravity (c.f. the discussion in Refs.~\cite{carlip2008,ref1a,ref1b,ref1c}), and concerns about the incompatibility of a variety of suggested theories with quantum mechanics~\cite{hu2014}.

Here we focus on efforts to understand the capacity of gravity to communicate information, and in essence attempt to determine rigorous bounds on the necessity of \emph{quantum} communication in gravity.  Previously, two of us suggested~\cite{kafri2013} that any theory of gravity that did not communicate at the quantum level would have observable consequences due to a necessary `amplifier' noise that classical, continuous variable communication channels must display.  In essence, that paper presents a witness and verify protocol, in which the potential quantum communication via gravity is constrained by measuring the linear response of two gravitationally coupled systems, with the guarantee that non-quantum theories for such interactions lead to a necessary noise rate that could be observed.  
We required that local experiments, of the type considered in quantum information and in particle physics, can observe quantum behavior, including violation of Bell's inequality.  We denote these requirement local operations and classical communication (LOCC), following the quantum information literature.  

However, that approach does not tackle any of the deep challenges any such `classical communication only' theory would provide when attempting to reproduce a wider class of dynamics, including the simple observation of a $1/r^2$ force law at low (non-relativistic) energies.  
Here we examine any theory of gravity with the following properites: its low energy theory  sector reproduces the expected linear response behavior between test masses due to low-curvature general relativity, i.e., Newton's gravitation law, and in that same sector, it does not enable quantum communication.  With this class of theories in mind, we combine extensions of our prior approach with quantum measurement and feedback techniques that we establish in Ref.~\cite{milburn2014}, which shows that measure-and-feedback saturates the classical noise bound of our general witness-and-verify approach.  In essence, our `classical communication only' model can be seen as an extension of measurement-based localization theories such as Ref.~\cite{diosiMaster}.
We show that these extensions provide a stringent bound on non-quantum theories of interacting particles that reproduce $1/r$-like interactions, as the requirement that dynamics generated by feedback arise from a measurement naturally balance the strength of the measurement to achieve the desired effective force `law'.  In other respects, the measurement component of our hypothetical measure-and-feedback gravity is similar in spirit to the holographic screen work of Ref.~\cite{verlinde2011}.  Crucially, this model naturally leads to a fundamental, observable noise term.

With this model, we can calculate the position-based dephasing of a single particle superposition and an associated heating in a massive object, reproducing key features of Ref.~\cite{diosiMaster}. Our results give a length scale for such `classical communication only' theories which is well in excess of $10^{-13}$ m.  Such long length scales (and correspondingly low energy scales) remain to be tested explicitly in the laboratory.  However, an intuitive picture of particle physics suggests that for an effective theory, such a cutoff (i.e., keV energies) would have observable consequences above that energy scale, specifically in the range of standard atomic physics. 
On the other hand, the extremely small coupling constant $G$ means that observable consequences may still have eluded experimental efforts, so we cannot preclude classical theories at this point.

\section{Measure-and-feedback interactions on a lattice}
Consider our hypothetical situation: gravitational interactions are observed to follow a $1/r$ behavior at low energies but are not allowed to transmit quantum information.  Within this context, we would like a consistent model for understanding the low-energy (Newtonian) limit for gravity.  For simplicity, we consider a length cutoff  $a$ below which our low-energy theory breaks down.  Given this setup, we can consider a range of possible classical communication only theories.  Fortunately, Ref.~\cite{milburn2014} allows us to consolidate all such theories to the noise bound by only considering a measure and feedback approach.  We regard this type of theory as an ``impostor'' theory of gravity, with no particularly good physical basis.  However, saturation of the noise bound indicates it is as good as any other classical communication only theory, and thus we use it for its conceptual and calculational simplicity.

\begin{figure}
\includegraphics[width=5.0in]{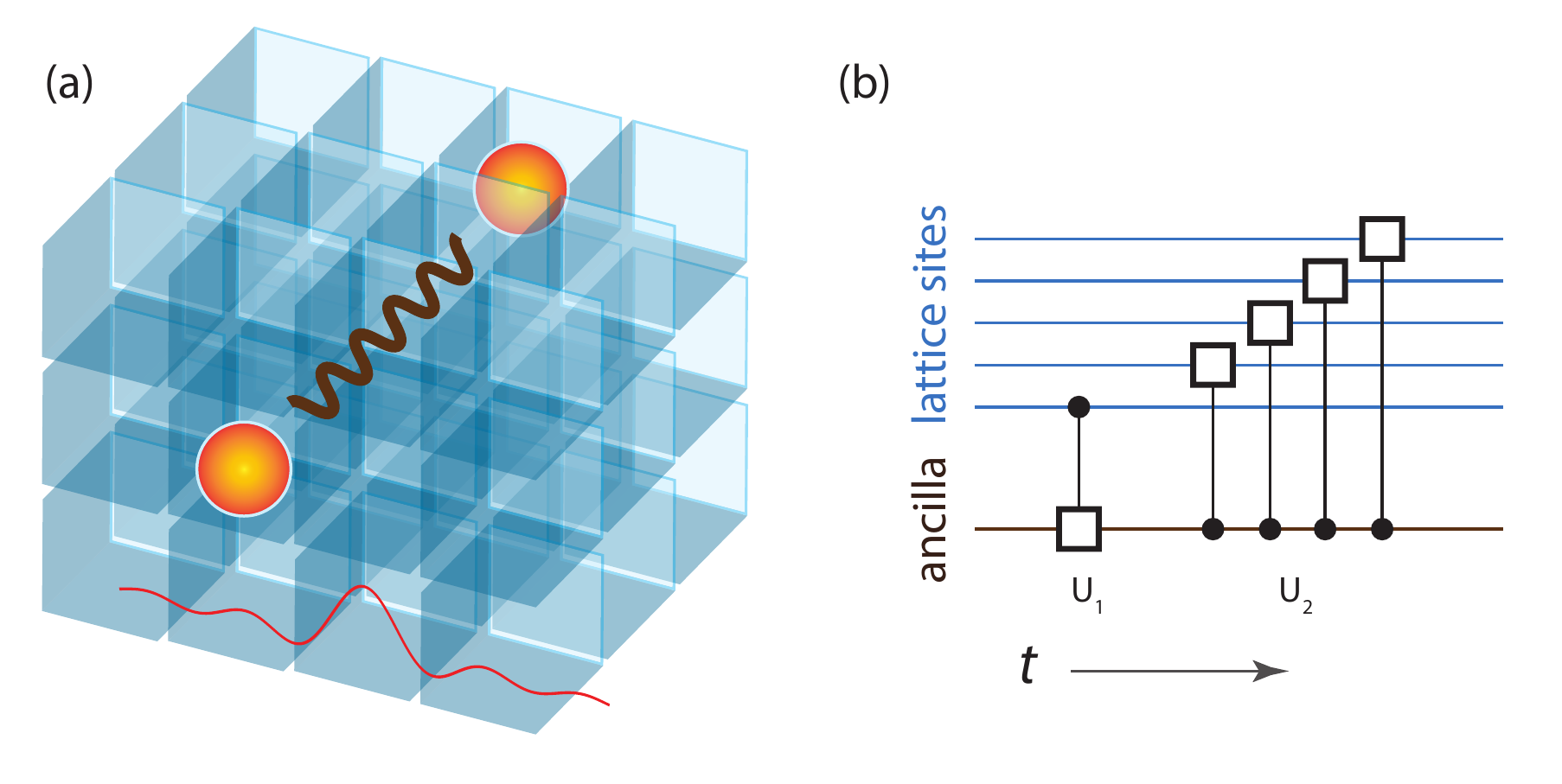}
\caption{
(a) Pictorial representation of particles on a lattice, with the lattice site mode function $f_j(x)$ plotted in red, and potential communication between particles indicated pictorially.
(b) Circuit for implementing one portion of the interaction, $V_j$, between a potential mass at lattice site $j$ and all other lattice sites, via an ancillary harmonic oscillator.  The ancilla is discarded at the end of each hypothetical implementation of this measure-and-feedback stage.
\label{fig1}
}
\end{figure}

Starting with single particles in one dimension, our length cutoff can be made exact by replacing the position basis $\ket{x}$ with a discrete variable representation corresponding to the projector $\hat \Pi = \int_{|p| < p_c} dp \ketbrad{p}$.  This is analogous to moving to a tight-binding-type model\footnote{We can use a different projector, for example the one that leads to a tight-binding model by summing over a countably infinite set of $k$'s, without much change of the overall picture.}.  Defining $p_c = \frac{\pi \hbar}{a}$ as a momentum cutoff in the single particle picture, this yields an integer-indexed set of states $\ket{x_j}$ with a position basis representation, in one dimension,
\begin{equation}
f_j(x) \equiv \langle x | x_j \rangle = \sqrt{\frac{\hbar}{\pi p_c}} \frac{\sin[p_c (x-x_j)/\hbar ]}{x-x_j}
\end{equation}
with $x_j = a j$, as shown pictorially in Fig.~\ref{fig1}a.   A similar picture was suggested in Ref.~\cite{ralph2009}.  

Looking in the second quantized picture, we work with a scalar field with mass $m$ and a field operator $\hat \psi(x)$.  This operator can be approximated to our length scale $a$ by transforming $x$ to $k$-space, then truncating, then transforming back:
\begin{eqnarray}
\hat a_j &=& \sqrt{\frac{\pi \hbar}{ p_c}} \int \frac{dq}{\sqrt{2\pi}} e^{-i q x_j}  \theta(p_c/\hbar - |q|) \hat \psi_q \nonumber \\
& =  &\sqrt{\frac{\pi \hbar}{ p_c}}  \int \frac{dq}{\sqrt{2\pi}} e^{-i q x_j}  \theta(p_c/\hbar - |q|) \int \frac{dx'}{\sqrt{2\pi}} e^{i q x'} \hat \psi(x') \nonumber \\
& = & \int f_j(x') \hat \psi(x') dx'
\end{eqnarray}
with the inversion relation for a Fourier transform on a finite domain
\[
\hat \psi_q = \sqrt{\frac{\hbar}{2 p_c}} \sum_j \hat a_j e^{i q x_j}\ 
\textrm{and} \ 
\Pi \hat \psi(x) \Pi = \sum_j f_j(x) \hat a_j .
\]
Regarding the commutation relations, we have for a bosonic field
\[
[ \hat a_j,\hat a_l^\dag ] = \int dx f(x-x_j) f(x-x_l) = \delta_{jl}
\]
More generally, these position-based field operators satisfy the same commutation relations as the continuum field, as expected.  Higher dimensions are straightforward to add, and we assume three spatial dimensions in the remainder of this discussion.  The free field Hamiltonian is
\[
\hat H_0 = \frac{\hbar^2}{2 m} \int d^3q  \theta(p_c/\hbar - |\bar q|) |\bar q|^2 \hat \psi_{\bar q}^\dag \hat \psi_{\bar q}
 \]
We can now re-express our expected gravitational interaction using these operators.  Specifically, starting with 
\[
\hat V = -\frac{G m^2}{2} \int d^3x d^3y \frac{\hat \psi^\dag(\bar x) \hat \psi^\dag(\bar y) \hat \psi(\bar y) \hat \psi(\bar x)}{|\bar x-\bar y|}
\]
where $m$ is the mass of our scalar field,
we project with $\Pi$.  This produces a sum $-\frac{G m^2}{2} \sum_{ijkl} I_{ijkl} \hat a_i^\dag \hat a_j^\dag \hat a_k \hat a_l$ with the integral $I_{ijkl} = \int d^3x d^3y \frac{f_i(x) f_j(y) f_k(y) f_l(x)}{|x-y|}$.  Looking at $I_{ijkl}$, we have three terms that contribute after projection with $\Pi$: a direct, Coulomb-like term ($I_{ijji}$, $i \neq j$) which contributes at long range,  exchange terms ($I_{iijj}$, $I_{ijij}$, and $I_{ijkl}$ with at most one index repeated) which decays over a length scale $\gtrsim a$ due to the reducing overlap of the sinc functions at long distance, and a self-interaction from the finite-size wave packet $I_{iiii}$ which we neglect as a background (constant) contribution for any particle conserving theory.

With these approximations, we have a softened Newtonian interaction
\begin{equation}
\hat \Pi \hat V\hat \Pi \approx \hat V_a \equiv -\frac{Gm^2}{2} \sum_{i \neq j} \hat a_i^\dag \hat a_j^\dag \hat a_i \hat a_j \frac{1}{|\bar x_i - \bar x_j| + a}  
\end{equation}
At this point, a Hamiltonian that contains both the kinetic terms $\hat H_0$ and interaction $\hat V$ provides a toy model for the quantum mechanics of interacting, gravitating particles.  We now consider what happens when we add the crucial constraint of this paper: that interactions through $\hat V_a$ do not convey quantum information.  More specifically, defining $\hat n_i = \hat a_i^\dag \hat a_i$ as a number operator measuring the local  particle number, we can emulate a set of pairwise interactions of the form $\hat V_{ij} = -\frac{G m^2}{2(|\bar x_i -\bar x_j |+ a)} \hat n_i \hat n_j$ with the minimum necessary back action using a measure-and-feedback approach with a (hypothetical) ancillary field.  In particular, the ancillary field stores a weak measurement result of the mass $m \hat n_i$ at a given location, which is then used to create a feedback force at distant locations.

Note that if we have multiple fields, we might write $\hat n_i$ as a weighted sum over occupation numbers of fields of different masses.  More general, relativistic scenarios for parameter estimation of a local Hamiltonian are also possible~\cite{downes2011}, and a simple analysis suggests that the $T_{00}$ component of the stress energy tensor is the more natural variable to observe in that regime. 
Here we are working strictly in the non-relativistic regime, and use $\hat n_i$ in what follows.

From the measure-and-feedback perspective, we expect for a protocol that mimics the effects of $\hat V_a$ necessarily acts via weak measurement of the set $\{ \hat n_j \}$.  Consequently, we anticipate an unavoidable noise in observables that do not commute with elements of this set.  
We can understand this unavoidable noise as arising due to local estimation of $\hat n_j$, i.e., a local measurement of the mass in a region of size $a^3$.

We now describe the measurement and feedback process illustrated in Fig.~\ref{fig1}b.  We recall that the interaction of interest is
\beq
\hat V_a = \hbar \sum_{j} \sum_{k \neq j} \chi_{jl} \hat n_j \hat n_k
\eeq
with $\chi_{jk} = - \frac{1}{2 \hbar} \frac{G m^2}{|\bar x_j - \bar x_k| + a}$.
For each position $j$ and at each time step $[t,t+\tau)$ we define an auxiliary harmonic oscillator with (dimensionless) canonical variables satisfying $[\hat X_j,\hat P_j] = i$.  Weak measurement with a strength $\xi$ corresponds to the unitary $\hat U_1 = \exp(-i \sqrt{\xi \tau} \hat n_j \hat P_j)$, where $\xi$ has units [s$^{-1}$].  This measurement is stored in $\hat X_j$, which can then induce a force via $\hat U_2 = \exp(-i  \sqrt{2 \tau/\xi} \hat X_j \sum_{k \neq j} \chi_{jk} \hat n_k )$. Using $\exp(r A)\exp(r B) = \exp(r A + r B + \frac{r^2}{2}[A,B] + O(r^3))$, we have that,
\[
\hat U_2 \hat U_1 = \exp\l(-i \sqrt{2\tau}( \sqrt{\xi}\hat n_j \hat P_j + \sqrt{1/\xi} \hat O_j \hat X_j) - i \tau \hat n_j \hat O_j+ O(\tau^{3/2})\r)
\]
where we have defined the feed-back operator,
$$\hat O_j =  \sum_{k \neq j} \chi_{jk} \hat n_k$$
That is, we have the desired interaction $-i \tau\hat n_j\sum_{k} \chi_{j k} \hat n_k$, as well as terms representing the strength of the measurement and feedback \footnote{As we show below, all terms of order $\sqrt{\tau}$ vanish in the reduced master equation of $\hat \rho$.}. Importantly, after each time step $\tau$, the auxiliary harmonic oscillator is discarded and replaced with another in the ground state, $\ket{\textrm{vac}}$.

The effective time evolution emerges in the limit $\tau \rightarrow 0$, so the master equation $\partial_\tau \hat \rho = \mc{L}(\hat \rho)$ is obtained by expanding $\hat U_2 \hat U_1 \hat \rho \otimes \ketbrad{\textrm{vac}} \hat U_1^\dagger \hat U_2^\dagger $ to order $\tau$ then tracing out the ancillary oscillator (this process is repeated at each site). Substituting from the above equation we compute,
\ba \nonumber
 && \l(\hat \rho  - i\tau[ \hat n_j \hat O_j, \hat \rho ]\r)\otimes \ketbrad{\textrm{vac}} \\
\nonumber
& -& i \sqrt{2\tau}\l[\sqrt{\xi}\hat n_j \hat P_j + \sqrt{1/\xi} \hat O_j \hat X_j, \hat \rho\otimes \ketbrad{\textrm{vac}}\r] \\
\nonumber &-& \tau \l[\sqrt{\xi}\hat n_j \hat P_j + \sqrt{1/\xi} \hat O_j \hat X_j,\l[\sqrt{\xi}\hat n_j \hat P_j + \sqrt{1/\xi} \hat O_j \hat X_j, \hat \rho\otimes \ketbrad{\textrm{vac}}\r]\r]
\ea
Taking the trace over the ancilla, the first line gives the expected interaction term $\hat \rho - i \tau [\hat n_j \hat O_j , \hat \rho]$, while the second line vanishes as $\bra{\textrm{vac}}\hat X_j \ket{\textrm{vac}} = \bra{\textrm{vac}}\hat P_j \ket{\textrm{vac}} = 0$. The final line is associated with the back-action and feedback noise introduced into the system. Using $\mean{\hat P_j^2} = \mean{\hat X_j^2} = 1/2$, $\mean{  \hat X_j \hat P_j +  \hat P_j\hat X_j} = 0$, and the cyclic property of the trace, one can show it is equal to, 
\beq
\mathcal{L}_{\textrm{noise}}^j  (\hat \rho) = -\frac{\xi}{2} [\hat n_j,[\hat n_j,\hat \rho]] -   \frac{1}{2 \xi} [\hat O_j,[\hat O_j,\hat \rho]] \label{e:L}
\eeq
The complete master equation is obtained by summing over effects from all sites $j$, as well as including local time evolution due to $\hat H_0$ for time $\tau$.

Equation~\ref{e:L} represents the central identity of this paper.  Choice of the measurement strength parameter $\xi$ may be optimized, as we describe below, to bound various potential non-quantum theories.  Curiously, this equation is reminiscent of the similar master equation deduced by Diosi~\cite{diosiMaster}, which was seen as necessary to maintain reversibility in the Newton-Schrodinger equation evolution.  Here we have explicitly constructed the theory to be compatible with standard quantum mechanics, and find that the intuitive picture elucidated by Diosi appears to be comparable to the measure-and-feedback approach. 


%


\section{Convergence of the noise operator}

Using this formalism, we now calculate the effects of the measurement and feedback noise super-operator, $ \mathcal{L}_{\mbox{noise}} = \sum_j \mc{L}_{\mbox{noise}}^j$. To do so we change to the Heisenberg picture, for which the relevant super-operator is its Hermitian adjoint, defined explicitly through the Hilbert-Schmidt inner product, $\tr{\mathcal{L}_{\mbox{noise}}^\dagger(\hat A^\dagger) \hat B} \equiv \tr{\hat A^\dagger \mathcal{L}_{\mbox{noise}}(\hat B)}$. By the cyclic property of the trace, we have that in this case the noise super-operator is self-adjoint, i.e. $\mathcal{L}_{\mbox{noise}}^\dagger = \mathcal{L}_{\mbox{noise}}$.

Before proceeding to calculating the noise and dephasing, we first must address a technical point. Strictly speaking, the action of $\mathcal{L}_{\mbox{noise}}^\dagger$ is not well defined on all operators, as the latter sum in Equation \ref{e:L} may diverge. In our analysis we therefore only consider operators which conserve the total particle number, since (as we show below) the action of $\mc{L}_{\mbox{noise}}^\dagger$ is bounded on these. In essence, we are restricting the class of physical operations and observables to those satisfying a total particle number super-selection rule (which the Hamiltonian and $\mc{L}_{\mbox{noise}}$ already satisfy). More specifically, we consider the dynamics of operators of the form,
\beq
\hat C = \hat a_{j_1}^\dagger \hat a_{j_2}^\dagger \,...\, \hat a_{j_N}^\dagger \hat a_{i_1} \hat a_{i_2} \,...\, \hat a_{i_N}
\eeq
for some positive integer $N$. We posit that any (possibly unbounded) operator which conserves total particle number $\l(\sum_k \hat n_k\r)$ and can be written in normal order can necessarily be written as sums of terms looking like $\hat C$. To consider bounded operators which conserve total particle number, we may equivalently assume $\hat C$ is of the form $ \hat a_{j_1}^\dagger \hat a_{j_2}^\dagger \,...\, \hat a_{j_N}^\dagger \ketbrad{\textrm{vac}}\hat a_{i_1} \hat a_{i_2} \,...\, \hat a_{i_N}$, where $\ket{\textrm{vac}}$ is the many-body vacuum state corresponding to zero mass. Since $[\hat n_i,\ketbrad{\textrm{vac}}]=0$ for all $i$, the following arguments are identical in either case.  

To begin, we note the following identity
$$
[\hat A,\hat B_1 \hat B_2 \,...\, \hat B_N] = [\hat A,\hat B_1](\hat B_2 \,...\, \hat B_N) + \hat B_1 [\hat A,\hat B_2] (\hat B_3 \,...\, \hat B_N) + \,...\, + (\hat B_1  \,...\, \hat B_{k-1}) [\hat A,\hat B_N]
$$
Given $[\hat n_k,\hat a_i] = - \delta_{k i} \hat a_i $ and $[\hat n_k,\hat a_j^\dagger] = \delta_{k j} \hat a_j^\dagger$, we then quickly compute,
$$[\hat n_k,[\hat n_{k'},\hat C]] = \l( \sum_{n = 1}^N (\delta_{k j_n}- \delta_{k i_n})  \r) \l( \sum_{m = 1}^N (\delta_{k' j_m}- \delta_{k' i_m})  \r)\hat C\,.$$
Hence the action of $\mc{L}_{\mbox{noise}}^\dagger$ on $\hat C$ is
\begin{eqnarray}
\mc{L}_{\mbox{noise}}^\dagger(\hat C) &= -\frac{1}{2} \sum_l \l[ \xi  \l( \sum_{n = 1}^N (\delta_{l j_n}- \delta_{l i_n})  \r)^2+ \r.
 \\
& \ \ \l. \frac{1}{\xi}\sum_{k k'} \chi_{l k}\chi_{l k'}\l( \sum_{n = 1}^N (\delta_{k j_n}- \delta_{k i_n})  \r) \l( \sum_{m = 1}^N (\delta_{k' j_m}- \delta_{k' i_m})  \r) \r] \hat C \nonumber 
\end{eqnarray}  
 
Considering the term proportional to $\xi$, observe that the inner sum is just $\#\{j_n = l \} - \#\{i_n = l \}$, i.e. the number of $j_n$ indices equal to $l$ minus the same for $i_n$. Hence the sum over $l$ is simply a (finite) combinatorial function of the indices,
\beq
\label{combisum}
- \xi \frac{1}{2} \sum_l (\#\{j_n = l \} - \#\{i_n = l \})^2
\eeq 

Likewise, the term proportional to $\frac{1}{\xi}$ is
\begin{eqnarray}
& -\frac{1}{2 \xi} \sum_l\l( \sum_{k k'} \chi_{l k}\chi_{l k'}\l( \sum_{n = 1}^N (\delta_{k j_n}- \delta_{k i_n})  \r) \l( \sum_{m = 1}^N (\delta_{k' j_m}- \delta_{k' i_m})  \r) \r) \nonumber \\
&  = -\frac{1}{2 \xi} \sum_l  \l( \sum_{n = 1}^N (\chi_{l j_n}- \chi_{l i_n})  \r) \l( \sum_{m = 1}^N (\chi_{l j_m}- \chi_{l i_m})  \r) \nonumber \\
& = -\frac{1}{2 \xi} \sum_{n = 1}^N \sum_{m = 1}^N \sum_l (\chi_{l j_n}- \chi_{l i_n})(\chi_{l j_m}- \chi_{l i_m})
\end{eqnarray}
where we have only interchanged orders of summation involving finite sums. 
Using the definition of $\chi_{i j}$, this sum can be rewritten as
\begin{eqnarray}
\label{chisum}
\l(\frac{G m^2}{2 \hbar}\r)^2\sum_l &\frac{(d_{j_n l}-d_{i_n l})(d_{j_m l}-d_{i_m l})}{(d_{j_n l} + a)(d_{j_m l} + a)(d_{i_n l} + a)(d_{i_m l} + a)}
\end{eqnarray} 
 where $d_{i j} = |\bar x_i - \bar x_j|$. 

We note that the above sum is finite: From the triangle inequality, $|d_{i j} - d_{k j}| \leq d_{i k}$, we can bound the absolute value of the numerator by $ d_{j_n i_n}d_{j_m i_m}$, which is independent  of $\bar x_l$.  Hence the sum above is absolutely convergent, and is bounded in magnitude by the sum,
$$  d_{j_n i_n}d_{j_m i_m} \l(\frac{G m^2}{2 \hbar}\r)^2\sum_l \frac{1}{(d_{j_n l} + a)(d_{j_m l} + a)(d_{i_n l} + a)(d_{i_m l} + a)}$$
which is finite in three dimensions.   


\section{Dephasing of a single particle mass superposition}
We now calculate the spatial dephasing rate of a single particle evolving under these dynamics, and find $\xi$ that minimizes such dephasing. In the first quantized picture, its density matrix $\hat \rho_s$ would be written as
\beq
\hat \rho_s = \sum_{i j} \rho_{i j} \ketbradt{i}{j}
\eeq 
with $\ket{j}$ being the single particle state at site $j$. This maps into the second quantized picture as
\beq
\hat \rho = \sum_{i j}\rho_{i j} a_i^\dagger \ketbrad{\textrm{vac}} a_j 
\eeq
where $\ket{\textrm{vac}}$ is the vacuum state. Defining the operators
\beq
\hat C_{j i} \equiv \hat a_j ^\dagger \ketbrad{\textrm{vac}} \hat a_i
\eeq
and noticing that $\rho_{i j} = \tr{\hat C_{j i} \hat \rho}$, we have that for dynamics solely determined by the operators in equation (7),
\beq
\dot \rho_{i j} =\tr{\hat C_{j i} \mathcal{L}_{\mbox{noise}}(\hat \rho)} = \tr{\mathcal{L}_{\mbox{noise}}^\dagger(\hat C_{j i}) \hat \rho}
\eeq

Finally, from Equations \ref{combisum} and \ref{chisum}, for $j \neq i$ we have,
\beq
\mc{L}_{\mbox{noise}}^\dagger(\hat C_{j i}) = -(\xi + \frac{1}{2 \xi} \kappa_{j i}^2) \hat C_{j i}
\eeq
where the latter frequency is 
\beq
\label{kappa}
\kappa_{j i}^2 = \l(\frac{G m^2}{2 a \hbar}\r)^2\sum_l \frac{(d_{i l}/a-d_{j l}/a)^2}{(d_{j l}/a + 1)^2(d_{i l}/a + 1)^2}
\eeq
The dephasing rate is minimized at $\xi^2 = \frac{1}{2} \kappa_{j i}^2$, giving a minimal decay rate of $\sqrt{2} \kappa_{j i}$. We now proceed to bound this rate from below.

Assuming that $|\mathcal{D}| = \frac{1}{ a}d_{j i} \gg 1$, the sum of Equation \ref{kappa} is approximated by the dimensionless integral expression,
\beq
\label{integral}
I = \int \mbox{d}^3u \frac{ (|u - \bar r_j| - |u - \bar r_i|)^2}{ (1 + |u - \bar r_j|)^2 (1 + |u - \bar r_i|)^2}
\eeq 
where $\bar r_j = \bar x_j/a$. By translating $u \rightarrow u + \frac{\bar r_i + \bar r_j}{2}$ and then rotating so that the vector $\bar \mathcal{D} =\bar r_j - \bar r_i $ is on the $z$ axis, this expression can be written in cylindrical coordinates as
\[
 2 \pi \int_0^\infty \mbox{d}r  \int_{-\infty}^\infty  \mbox{d}z \frac{ r (\sqrt{r^2 + (z-|\mathcal{D}|/2)^2} - \sqrt{r^2 + (z+|\mathcal{D}|/2)^2})^2 }{ (1 + \sqrt{r^2 + (z-|\mathcal{D}|/2)^2})^2 (1 + \sqrt{r^2 + (z+|\mathcal{D}|/2)^2})^2} 
\]
It is then straightforward to show that the numerator is larger than $r ( z |\mathcal{D}|)^2/ (z^2 + r^2 + |\mathcal{D}|^2/4)$, and the denominator is smaller than $(1 + \sqrt{r^2 + z^2 + |\mathcal{D}|^2/4})^4$ 
. The integral $I$ is therefore larger than
\begin{eqnarray*}
I > \int_0^\infty \mbox{d}r  \int_{-\infty}^\infty  \mbox{d}z & \frac{ 2 \pi (z |\mathcal{D}|)^2 r}{\left(1 + \sqrt{r^2 + z^2 + |\mathcal{D}|^2/4}\right)^6} =\frac{\pi^2}{2} |\mathcal{D}|(1 + O\left(|\mathcal{D}|^{-1}\right))
\end{eqnarray*}
Hence for a particle of mass $m$ in a spatial superposition of distance $d_{i j} = |\mathcal{D}| a$, we should observe a spatial dephasing rate on the order of at least
$\frac{G m^2}{2 a \hbar} \sqrt{\frac{d_{i j}}{a}}$.  We remark that this result is similar, at short distances, to the result of Ref.~\cite{diosi2007}, where the decoherence rate for a particle separated by one distance of `measurement' uncertainty goes as $\frac{GM^2}{2a\hbar}$.

We can already constrain such dephasing by considering interference of large molecules as described in Refs.~\cite{arndtNatComm2011,arndtRMP}. To do so, we treat the center of mass of each molecule as the effective position coordinate of the particles in our theory. These effective particles do not occupy the full mass distribution of the molecules, but rather are spread over a length on the order of their de Broglie wave length $(\hbar/p \sim 10^{-12}\mbox{ m}$.) This treatment is valid if, over the timescale of the experiment, this coordinate does not strongly interact with other (internal or environmental) degrees of freedom. Our assumption is supported, {\it a fortiori}, from the fact that interference is observed in the experiment. For a mass of $m\sim 5,000$ amu, a spatial superposition $\Delta \sim 0.5\ \mu$m at the second diffraction grating (set by the standing wave length scale), and an observed coherence time $\sim  1$ ms, we find that $a > 10^{-19}$ m.  This is 16 orders of magnitude larger than the Planck length, or about 100 GeV, a substantially lower energy scale than expected for gravitational physics to emerge. 

\section{Noise-induced heating}

Having established that restricting potential communication via gravity to being classical would cause a mass superposition to dephase, we next consider what happens in the case of
a large object composed of many particles. Although a detailed description requires the addition of other forces and particle species, for simplicity we focus on the consequences of including $\mc{L}_{\mbox{noise}}$ of equation \ref{e:L} in the general equation of motion. 
 Towards this end we will consider the evolution of the particles' momentum, which requires understanding how `hopping' terms in the discrete variable representation evolve under the super-operator $\mc{L}_{\mbox{noise}}$.

To start, we look at the evolution of $\hat a_j^\dagger \hat a_i$, for $j \neq i$: 
\beq
\mc{L}_{\mbox{noise}}(\hat a_j^\dagger \hat a_i) = -\l(\xi + \frac{1}{2 \xi}\kappa_{j i}^2
\r)\hat a_j^\dagger \hat a_i
\eeq 
where the latter constant is defined as in Equation \ref{kappa}. 
As before, since $\xi + \frac{1}{2 \xi} \kappa_{j i}^2  \geq \sqrt{2} \kappa_{j i}$ for all $\xi$, we have that for any $i \neq j$, the evolution of $\hat a_j^\dagger \hat a_i$ is damped at a rate on the order of at least $\frac{G m^2}{\hbar a}(1 + O( \sqrt{\frac{|d_{i j}|}{a}}))$. We now consider a finite region of space $S$, having a total mass operator 
$$\hat M_S = m \sum_{j \in S} \hat a_j^\dagger \hat a_j$$
Letting $\hat \Pi_S$ denote the projector onto the indices $j \in S$, the net momentum in this space is
\ba
 \bar p_S & =  \hat \Pi_S \l(a^3 \hbar \int  \mbox{d}^3k\, \bar k \, \tilde a_{\bar k}^\dagger \tilde a_{ \bar k}\r) \hat \Pi_S \\\
& = a^3 \hbar \sum_{j,j'\in S} \,  \hat a_{j'}^\dagger \hat a_{ j } \int  \mbox{d}^3k\, \bar k  e^{-i  (\bar x_j - \bar x_{j'}) \cdot \bar k}
\ea
where the integral is over $\bar k \in [-\pi/a,\pi/a]^{\otimes 3}$, and we used the defintion
$$\tilde a_{\bar k} = \frac{1}{\sqrt{(2 \pi)^3 }}  \sum_{j} e^{-i  \bar x_j \cdot \bar k} \hat a_j  $$
One may verify that the vector operator $\bar p_S$ satisfies the commutation relations $[\hat x_{S}^{(m)},\hat p_S^{(n)}] = i \delta_{m n} \hbar \sum_{j\in S} \hat a_j^\dagger \hat a_j$, where $x^{(m)}_{S} = \sum_{j\in S} x^{(m)}_j \hat a^\dagger_j \hat a_j$ is proportional to the $m$th coordinate of the center of mass position within region $S$. From the above analysis, we see that 
$\mc{L}_{\mbox{noise}}$ causes $\hat a_{j'}^\dagger \hat a_j$ to decays exponentially to 0 
at a rate at least $\sim \frac{G m^2}{\hbar a}$, hence, $\bar p_S$ tends to $0$ at this rate as well.  

Likewise, at a similar rate $\sim \frac{G m^2}{\hbar a}$ the variance of the net momentum approaches
\ba
\hat p_S^2 & =  \hat \Pi_S \l(\hbar^2 a^3 \int  \mbox{d}^3k\, |\bar k|^2\, \tilde a_{\bar k}^\dagger \tilde a_{\bar k}  \r)\hat \Pi_S \nonumber \\
& = \hbar^2 a^3 \sum_{j,j'\in S} \l(\int  \mbox{d}^3k e^{-i  (\bar x_j-\bar x_j') \cdot \bar k}\, |\bar k|^2\r)\, \hat a_{ j'}^\dagger \hat a_{ j}  \nonumber \\
& \rightarrow \hbar^2 \frac{a^3}{(2 \pi)^3} \int \mbox{d}^3k\, |\bar k|^2 \sum_{j\in S} \hat a_j^\dagger \hat a_j\\
&\nonumber = \frac{1}{m}\l(\frac{2 \pi \hbar}{a}\r)^2  \hat M_S
\ea
Since this asymptotic value is arbitrarily large as $a \rightarrow 0$, we can use it to approximate a heating rate of the space $S$ as
\ba
\mc{L}_{\mbox{noise}}\l(\frac{\hat p_S^2}{2 m} \r)& \sim \l(\frac{G m^2}{\hbar a}\r) \frac{1}{2 m} \l(\frac{1}{ m}\l(\frac{2 \pi \hbar}{a}\r)^2  \hat M_S \r)\nonumber\\
&\sim \frac{G \hbar}{a^3} \hat M_S
\ea
For an arbitrary region of space $S$, our theory therefore predicts a generic heating rate proportional to the the total mass contained in that space. 

We can use this heating rate, a consequence of restricting the gravitational potential to only convey classical information, to get experimental bounds for the length scale $a$ of our theory. Namely, a critical temperature $\sim 500$ nK and typical lifetime of 5 seconds for Rubidium-87 BEC's yields a per-atom heating rate of $\kappa_{BEC}\sim 10^{-30}$ J/s . Comparing this to the above bound with $\mean{\hat M_S} = M_{Rb} \sim 10^{-25}$ kg yields $a \gg \l(G M_{Rb} \hbar/\kappa_{Rb} \r)^{1/3} \sim 10^{-13}\mbox{ m}$. At a much different mass scale, we can also compare the predicted heating rate of the Earth to its total absorbed radiative energy from the Sun, on the order of $\sim 1 \times 10^{17}$J/s. Using $\mean{\hat M_S} = M_E$ gives a slightly tighter bound $a \gg 10^{-12}$m.

\section{Conclusion}
We have shown that combining observational constraints of interactions indicative of a $1/r^2$ force law with the concept that such forces cannot communicate quantum information leads to a simple model for `classical communication only' theories of gravity.  Our particular approach, using measurement and feedback as calculational tools, saturates the minimum noise and dephasing rates anticipated for any such model.  Examination of two particular cases -- matter wave interference and heating of massive objects -- indicate that most likely any `classical communication only' model of gravity breaks down at length scales larger than expected for new gravitational physics to emerge.  
However, stringent tests of softening of Newtonian interactions at short length scales remain before these bounds become complete.  Further extensions of these concepts to relativistic domains are in principle straightforward, and may provide a framework within which one can understand a wide variety of different heuristic models of gravitational decoherence.

The authors would like to thank J. Preskill, C. Caves, B.-L. Hu, and S. Schlamminger for helpful commentary and discussions on these topics, and the hospitality of the Kavli Institute for Theoretical Physics, where these ideas were first considered.  Support is provided by the NSF-funded physical frontier center at the JQI.


\begin{thebibliography}{10}
\bibitem{Karolyhazy}F. Karolyhazy, Il Nuovo Cimento A, {\bf 42}, 390 (1996).

\bibitem{Penrose}R. Penrose, Gen. Rel. Grav. {\bf 28} 581 (1996).

    \bibitem{Diosi}L. Di\'osi, Phys. Rev. A {\bf  40}, 1165 (1989); 
    
    \bibitem{diosi2007}
L. Di\'osi,
     J. Phys. A 40, 2989-2995 (2007).

    \bibitem{Pullin}R. Gambini, R. Porto, J. Pullin, New J Physics, {\bf  6} 45 (2004).

     \bibitem{Romero-Isart}O. Romero-Isart, A. C. Pflanzer, F. Blaser, R. Kaltenbaek, N. Kiesel, M. Aspelmeyer, J. I. Cirac, Phys. Rev. Lett. 107, 020405 (2011); 
    Oriol Romero-Isart, Phys. Rev. A 84, 052121 (2011).

     \bibitem{Yang}H. Yang, H. Miao, D.-S. Lee, B. Helou, and Y. Chen, Phys. Rev. Lett. 110, 170401 (2013).



\bibitem{carlip2008}
S. Carlip, Class. Quantum Grav. 25, 154010 (2008).

\bibitem{ref1a}
D. Giulini, A. Grossardt,
Class. Quantum Grav. 28, 195026 (2011).

\bibitem{ref1b}
D. Giulini, A. Grossardt,
Class. Quantum Grav. 29, 215010 (2012).

\bibitem{ref1c}
D. Giulini, A. Grossardt,
 	New J. Phys. 16, 075005 (2014).
 	
\bibitem{hu2014}
B. L. Hu,  Journal of Physics: Conference Series,  504, 012021 (2014).

\bibitem{kafri2013}
    D. Kafri, J. M. Taylor, arXiv:1311.4558 (2013).

\bibitem{milburn2014}
D. Kafri, J. M. Taylor, G. J. Milburn,  arXiv:1401.0946 (2014).

\bibitem{diosiMaster}
L. Di\'osi
Phys. Lett. 120A, 377-381 (1987); 

\bibitem{verlinde2011}
E. Verlinde, J. High Energy Phys. JHEP04(2011)029

\bibitem{ralph2009}
T. C. Ralph, G. J. Milburn, T. Downes, Phys. Rev A79 , 022121 (2009).

\bibitem{downes2011}
T. G. Downes, G. J. Milburn, C. M. Caves, http://arxiv.org/abs/1108.5220 (2011).




     
     \bibitem{arndtNatComm2011}
    Stefan Gerlich,	
    Sandra Eibenberger,	
    Mathias Tomandl,	
    Stefan Nimmrichter,	
    Klaus Hornberger,	
    Paul J. Fagan,	
    Jens T\"uxen,	
    Marcel Mayor	,
   \& Markus Arndt,
   Nat. Comm. 2, 263 (2011).

\bibitem{arndtRMP} Klaus Hornberger, Stefan Gerlich, Philipp Haslinger, Stefan Nimmrichter, and Markus Arndt,
Rev. Mod. Phys. 84, 157 (2012).

\bibitem{Streed2006}
Erik~W. Streed, Ananth~P. Chikkatur, Todd~L. Gustavson, Micah Boyd, Yoshio
  Torii, Dominik Schneble, Gretchen~K. Campbell, David~E. Pritchard, and
  Wolfgang Ketterle.
\newblock Large atom number bose-einstein condensate machines.
\newblock {\em Review of Scientific Instruments}, 77(2):--, 2006.




\end{thebibliography}
\end{document}